\documentstyle[12pt]{article}
\input html.sty
\input epsf
\begin{document}
\title{MATTERS OF GRAVITY, The newsletter of the APS Topical Group on 
Gravitation}
\begin{center}
{ \Large {\bf MATTERS OF GRAVITY}}\\
\bigskip
\hrule
\medskip
{The newsletter of the Topical Group on Gravitation of the American Physical 
Society}\\
\medskip
{\bf Number 16 \hfill Fall 2000}
\end{center}
\begin{flushleft}

\tableofcontents
\vfill
\section*{\noindent  Editor\hfill}

\medskip
Jorge Pullin\\
\smallskip
Center for Gravitational Physics and Geometry\\
The Pennsylvania State University\\
University Park, PA 16802-6300\\
Fax: (814)863-9608\\
Phone (814)863-9597\\
Internet: 
\htmladdnormallink{\protect {\tt{pullin@phys.psu.edu}}}
{mailto:pullin@phys.psu.edu}\\
WWW: \htmladdnormallink{\protect {\tt{http://www.phys.psu.edu/\~{}pullin}}}
{http://www.phys.psu.edu/\~{}pullin}\\
\hfill ISSN: 1527-3431
\begin{rawhtml}
<P>
<BR><HR><P>
\end{rawhtml}
\end{flushleft}
\pagebreak
\section*{Editorial}

Not much to report here. The newsletter is a bit late (it was due September
1st) due to last minute updates (TAMA and TOCO). I felt like a real editor
for a little while! The next newsletter is due February 1st.
If everything goes well this newsletter should be available in the
gr-qc Los Alamos archives under number gr-qc/0009060. To retrieve it
send email to
\htmladdnormallink{gr-qc@xxx.lanl.gov}{mailto:gr-qc@xxx.lanl.gov}
with Subject: get 0009060
(numbers 2-15 are also available in gr-qc). All issues are available in the
WWW:\\\htmladdnormallink{\protect {\tt{http://gravity.phys.psu.edu/mog.html}}}
{http://gravity.phys.psu.edu/mog.html}\\ 
A hardcopy of the newsletter is
distributed free of charge to the members of the APS
Topical Group on Gravitation upon request (the default distribution form is
via the web) to the secretary of the Topical Group. 
It is considered a lack of etiquette to
ask me to mail you hard copies of the newsletter unless you have
exhausted all your resources to get your copy otherwise.
\par
If you have comments/questions/complaints about the newsletter email
me. Have fun.
\bigbreak

\hfill Jorge Pullin\vspace{-0.8cm}
\section*{Correspondents}
\begin{itemize}
\item John Friedman and Kip Thorne: Relativistic Astrophysics,
\item Raymond Laflamme: Quantum Cosmology and Related Topics
\item Gary Horowitz: Interface with Mathematical High Energy Physics and
String Theory
\item Richard Isaacson: News from NSF
\item Richard Matzner: Numerical Relativity
\item Abhay Ashtekar and Ted Newman: Mathematical Relativity
\item Bernie Schutz: News From Europe
\item Lee Smolin: Quantum Gravity
\item Cliff Will: Confrontation of Theory with Experiment
\item Peter Bender: Space Experiments
\item Riley Newman: Laboratory Experiments
\item Warren Johnson: Resonant Mass Gravitational Wave Detectors
\item Stan Whitcomb: LIGO Project
\end{itemize}
\vfill
\pagebreak

\section*{\centerline {
Cosmic microwave background anisotropies:} \\
\centerline{tantalizingly close to expectations}}
\addtocontents{toc}{\protect\medskip}
\addtocontents{toc}{\bf Research Briefs:}
\addtocontents{toc}{\protect\medskip}
\addcontentsline{toc}{subsubsection}{\it  
Cosmic microwave background anisotropy experiments, by Sean Carroll}
\begin{center}
    Sean Carroll, University of Chicago\\
\htmladdnormallink{carroll@theory.uchicago.edu}
{mailto:carroll@theory.uchicago.edu},
\htmladdnormallink{
http://pancake.uchicago.edu/{\~{}}carroll/}
{http://pancake.uchicago.edu/~carroll/}
\end{center}

Ever since the detection of temperature anisotropies in the 
cosmic microwave background (CMB) by the COBE satellite in 1992,
cosmologists have anticipated that information about the amplitude
of these fluctuations across a range of angular scales could be an
extraordinarily powerful constraint on cosmological models
(see for example [1]).  Now a series of new
experiments --- the TOCO98 run of the MAT ground-based
telescope in Chile [2], the 
balloon-borne Boomerang experiment flown both in Texas
[3] and Antarctica [4], and the balloon-borne
Maxima experiment flown in Texas [5]
--- have turned these expectations into reality.

The figure shows the new results combined with
previous experiments,
presented as amplitude of fluctuation vs.\ multipole moment $l$
in a spherical harmonic decomposition.
\begin{figure}[ht]
  \epsfxsize = 10cm
  \centerline{\epsfbox{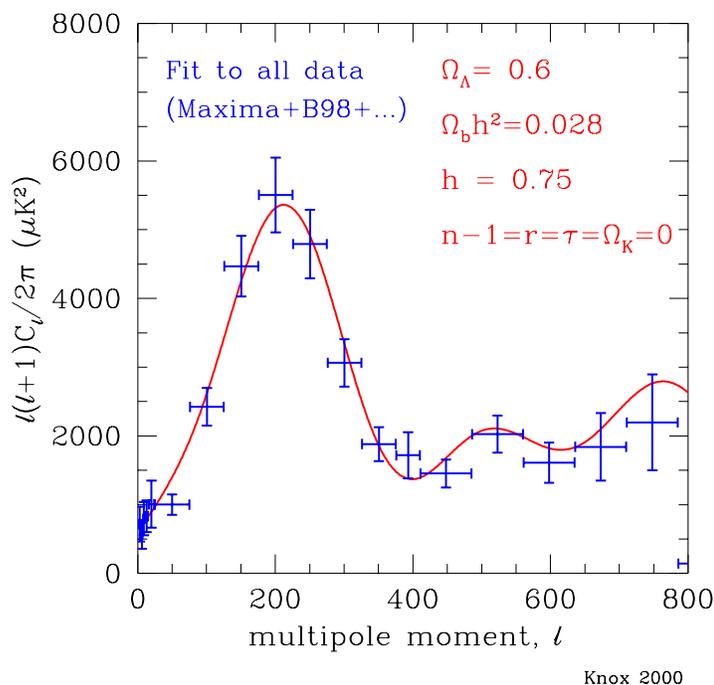}}
  \caption{Amplitude of CMB temperature anisotropies, as a function
  of multipole moment $l$ (so that angular scale decreases from
  left to right).  The data points are averaged from all of the
  experiments performed as of Summer 2000.  The curve is a theoretical
  model with scale-free adiabatic scalar perturbations in a flat
  universe dominated by a cosmological constant, with a slightly
  higher baryon density than implied by big-bang nucleosynthesis.
  Courtesy of Lloyd Knox.} 
  \label{knoxcmb}
\end{figure}
Angular size decreases from left to right in the figure, as
multipole moment is related to angular scale roughly by 
$\theta \sim 180^\circ /l$.
This plot manifests three crucial features:
\begin{enumerate}
\item A well-defined, narrow ($\Delta l/l \sim 1$) peak in the
power spectrum.  This is strong evidence in favor of ``inflationary''
primordial perturbations.
\item Location of the peak at $l\sim 200$.  This is strong evidence
in favor of a nearly flat spatial geometry ($\Omega_{\rm tot} = 1$).
\item A secondary peak ($l\sim 500$) which is rather small, if one
is indicated at all.  Previous best-fit models predicted a noticeable
peak at this location; this might be evidence of tilt in the 
perturbation spectrum, a higher-than-expected baryon density, or
more profound physics.
\end{enumerate}
Let's consider each of these features in turn.

The adjective ``inflationary'' refers to adiabatic perturbations
that have been imprinted with (nearly) equal amplitudes on all 
scales (both greater and less than the Hubble radius $H^{-1}$)
before recombination.  ``Adiabatic'' means that fluctuations in
each species are correlated, so that the number density ratios
of photons/baryons/dark matter are spatially constant.  (Here,
``baryons'' is cosmology-speak for ``charged particles.'')
These are the kinds of perturbations predicted by inflationary
cosmology; it is entirely possible that a mechanism other than inflation
could generate perturbations of this type, although no theories
which do so have thus far been proposed.  When we observe
temperature fluctuations in the CMB, on scales which are larger
than the Hubble radius at recombination the dominant effect is
the gravitational redshift/blueshift as photons move through 
potential wells (the Sachs-Wolfe effect), while on smaller scales
the intrinsic temperature anisotropy is dominant.  An adiabatic mode
of wavelength $\lambda$ (which grows along with the cosmic scale
factor) will remain approximately constant in amplitude while
$\lambda > H^{-1}$, after which it will begin to evolve
under the competing effects of self-gravity (which works to 
increase the density contrast) and radiation pressure (which works
to smooth it out).  The result is an acoustic wave which
oscillates during the period between when the mode becomes
sub-Hubble-sized and recombination (when radiation pressure
effectively ends).  As the wave evolves it is also damped as
photons dissipate from overdense to underdense regions.  We therefore
expect to see a series of peaks in the CMB spectrum, with the largest
peak corresponding to a physical length scale equal to that of the
Hubble radius at recombination.  A crucial point is that the
sharpness of this peak is evidence for the temporal coherence of
the waves --- the evolution of a wave at any one wavelength is
related in a simple way to that at other wavelengths, which enables
the spectral features to be well-defined (see [6] for
a discussion).  In models where the
perturbations are continually generated at all times (such as with
topological defects), or models of ``isocurvature'' fluctuations
in which different species are uncorrelated,
this coherence is absent, and it is very
difficult to get a sharp peak.  The new observations thus
strongly favor primordial adiabatic perturbations.

As mentioned, the location of the first peak corresponds to the
Hubble radius at the last scattering surface,
$H_{\rm LS}^{-1}$.  In a spatially flat universe,
the observed angular scale of the peak would be the ratio of
$H_{\rm LS}^{-1}$ to the angular diameter distance 
$r_\theta$ between us and
the surface of last scattering.  It turns out that, in a
Friedmann-Robertson-Walker cosmology with plausible values of the
various cosmological parameters, both $H_{\rm LS}^{-1}$ and $r_\theta$
depend on these parameters in roughly the same way: they are
each proportional to $H_0^{-1}/\sqrt{\Omega_{{\rm M}0}}$, where
$\Omega_{\rm M}$ is the ratio of the matter density to the critical
density and subscripts $0$ refer to quantities evaluated at the
present time.  The ratio $H_{\rm LS}^{-1}/r_\theta$ 
is thus approximately independent of the cosmological parameters.  
The observed angular scale of the first peak therefore depends 
primarily on the spatial geometry through which
the photons have traveled; in a positively/negatively curved
space, a fixed physical size corresponds to a larger/smaller
angular size.  The spatial geometry can be quantified by the total
density parameter $\Omega_{\rm tot}$, and the angular dependence
of the peak turns out to be $l_{\rm peak}\sim 200
\Omega_{\rm tot}^{-1/2}$.  Thus, the observed peak at $l\sim 200$
provides excellent evidence for a flat universe.  The most recent
data are sufficiently precise that sub-dominant
effects become relevant, and more careful analyses are necessary
[7].  The quantitative results depend somewhat on 
which parameters are allowed to vary and which additional
data are taken into account; the CMB data alone
are actually best fit by a
universe with a very small positive spatial curvature, but a
perfectly flat universe is within the errors, while an open
matter-dominated universe with $\Omega_{\rm tot}<0.5$ is strongly
ruled out.  Taking existing data on the Hubble parameter and
large-scale structure distribution into acount implies the need
for a positive cosmological constant, thus providing some 
independent confirmation for the striking supernova results
[8].

The most unexpected feature of the observed CMB power spectrum,
from the point of view of previously favored cosmological parameters,
is the absence of an easily distinguishable secondary peak.
It turns out that the expected peak can be suppressed in two
straightforward ways: by ``tilting'' the primordial spectrum so that 
there is slightly less power on small scales, or by increasing the
baryon-to-photon ratio.  The tilting option, while plausible, is
hard to accommodate within simple inflationary models, as a 
sufficient tilt is necessarily accompanied by additional tensor
fluctuations on large scales [9], ruining the rest
of the fit.  The baryon density is most conveniently expressed
in terms of $\Omega_{\rm b}h^2$, where $\Omega_{\rm b}$ is the 
density parameter in baryons and $h = H_0/(100$~km/sec/Mpc).
The CMB data [7] imply $\Omega_{\rm b}h^2 = 0.032\pm 0.009$, 
while big-bang nucleosynthesis [10] implies 
$\Omega_{\rm b}h^2 = 0.019 \pm 0.002$ (at $95\%$ confidence),
with an ``extreme upper limit'' [11] of 
$\Omega_{\rm b}h^2 \leq 0.025$.  Hence, consistency is just barely
preserved at the edges of the allowed values.
It would seem at this point most likely that some combination
of slight tilt and ordinary experimental error have combined to
create this apparent tension, but there remains the possibility of
interesting new physics.  (Note that the {\it upper} limit on the
baryon density provides additional support for the necessity
of non-baryonic dark matter.) 

The new CMB data are in a sense the idea experimental result, in
that they provide useful constraints within the context of a 
successful theory while raising questions about aspects of that
theory that can only be addressed by future experiments.  The
near future will see a number of new balloon, ground-based and
satellite measurements of the CMB power spectrum on even smaller
angular scales (including the presumed location of the third
peak and beyond), which should reveal whether we are seeing a
spectacular confirmation of the standard cosmology or the first
signs of important deviations from it.

{\bf References:}

[1]
G.~Jungman, M.~Kamionkowski, A.~Kosowsky and D.~N.~Spergel,
Phys.\ Rev.\  {\bf D54}, 1332 (1996)
[astro-ph/9512139];
W.~Hu, N.~Sugiyama and J.~Silk,
Nature {\bf 386}, 37 (1997)
[astro-ph/9604166].

[2]
A.~D.~Miller {\it et al.},
Astrophys.\ J.\  {\bf 524}, L1 (1999)
[astro-ph/9906421].

[3]
A.~Melchiorri {\it et al.},
Astrophys.\ J.\  {\bf 536}, L63 (2000)
[astro-ph/9911445].

[4]
P.~de Bernardis {\it et al.},
Nature {\bf 404}, 955 (2000)
[astro-ph/0004404].

[5]
S.~ Hanany  {\it et al.},
astro-ph/0005123.

[6]
L.~Knox, astro-ph/0002163.

[7]
A.~E.~Lange {\it et al.},
astro-ph/0005004;
A.~Balbi {\it et al.},
astro-ph/0005124;
A.H.~Jaffe {\it et al.},
astro-ph/0007333;
M.~White, D.~Scott and E.~Pierpaoli,
astro-ph/0004385;
M.~Tegmark, M.~Zaldarriaga and A.~J.~Hamilton,
astro-ph/0008167.

[8]
A.G. Riess {\it et al.}, {\sl Astronomical J.}
{\bf 116}, 1009 (1998) [astro-ph/9805201];
S. Perlmutter {\it et al.}, {\sl Astrophys. J.}
(1999) [astro-ph/9812133];
S.~M.~Carroll,
astro-ph/0004075.

[9]
A.~R.~Liddle and D.~H.~Lyth,
Phys.\ Lett.\  {\bf B291}, 391 (1992)
[astro-ph/9208007].

[10]
K.~A.~Olive, G.~Steigman and T.~P.~Walker,
Phys.\ Rept.\  {\bf 333-334}, 389 (2000)
[astro-ph/9905320];
S. Burles, K.~M.~Nollett, J.~N.~Truran and M.~S.~Turner,
Phys.\ Rev.\ Lett.\  {\bf 82}, 4176 (1999)
[astro-ph/9901157];
D. Tytler, J.M. O'Meara, N. Suzuki and D. Lubin,
astro-ph/0001318.

[11]
S. Burles, K.~M.~Nollett and M.~S.~Turner,
astro-ph/0008495.

\vfill
\pagebreak
\parskip=5pt
\section*{\centerline {
LISA Project Update}}
\addcontentsline{toc}{subsubsection}{\it  
LISA Project Update by Bill Folkner}
\begin{center}
    W. M. Folkner, Jet Propulsion Laboratory, Caltech\\
\htmladdnormallink{william.folkner@jpl.nasa.gov}
{mailto:william.folkner@jpl.nasa.gov}
\end{center}

        The last 18 months have been very important for the Laser
Interferometer Space Antenna project.  Several advisory panels have
made very positive recommendations regarding the prospects for a
gravitational-wave observatory in space.  NASA has begun serious
planning for funding the LISA project and associated technology
development starting in 2002 while continuing mission studies.  The
European Space Agency has continued technology development in key
areas and funded a Phase A study under a contract led by Dornier
Satellite Systems (now part of Astrium).  Informal discussions on a
partnership arrangement between NASA and ESA have continued and are
about to transition to a more formal working arrangement.  The
prospects for achieving a launch of LISA towards the end of the decade
are very good.  

In 1999 NASA Office of Space Science conducted a
series of meetings for the purpose of updating its Strategic Plan.
(These updates are held approximately every three years.) A key
meeting of the Structure and Evolution of the Universe Subcommittee
was held in February 1999 to consider future missions within the SEU
theme.  Following that meeting the SEUS recommend that LISA be
included in the 2000 OSS Strategic Plan as a candidate for a New Start
in the 2005-2008 time frame.  The SEUS recommendations are published
in the SEU Roadmap 
(\htmladdnormallink{http://universe.gsfc.nasa.gov/roadmap.html}
{http://universe.gsfc.nasa.gov/roadmap.html}
).  The
recommendation of the SEUS was forwarded for consideration by the
NASA's Space Science Advisory Committee.  At a meeting held in
November 1999 (in Galveston) the SSAC endorsed the SEUS recommendation
of LISA for inclusion in the 2000 OSS Strategic Plan, which will be
reflected in the Plan when published in the upcoming months.  Outside
NASA two panels commissioned by the National Research Council have
given LISA high priority for future missions.  The Committee on
Gravitational Physics, chaired by Jim Hartle, gave a low-frequency
gravitational-wave observatory its highest priority for space missions
in gravitational physics (\htmladdnormallink{
http://www.nap.edu/books/0309066352/html/}
{http://www.nap.edu/books/0309066352/html/}).
The Astronomy and Astrophysics Survey Committee, chaired by Joe Taylor
and Chris McKee, ranked LISA as second highest priority among
moderate-scale projects for the coming decade 
(\htmladdnormallink{
http://books.nap.edu/catalog/9839.html}
{http://books.nap.edu/catalog/9839.html}
).  

NASA funding for the
implementation of  moderate missions has historically been provided
by specific congressional line items in the NASA budget for each
project.  More recently congress has approved several continuing line
items for multiple missions (e.g. Origins, Mars Exploration).  NASA
has decided to request a new continuing line item, called Cosmic
Journeys, for the SEU theme to cover multiple future missions
including LISA.  If approved, significant advanced technology
development for LISA would begin in 2002 with a nominal launch date in
2010.  

Prior to approval for construction of the mission, the
technology needed to achieve the science goals must be demonstrated to
a suitable level.  For Technology Plan for LISA has been developed and
reviewed by an advisory group.  One of the key technology issues for
LISA is for the test masses to be free of unwanted forces that would
cause motions larger than those cause by gravitational waves.  For the
frequency and sensitivity desired for LISA the disturbances are
required to be less than $10^{-16} G$ for times scales of 100 to 10,000
seconds.  This level of performance seems achievable, and detailed
calculations of expected forces for some designs indicate that the
requirements can be met.  But the required performance is far beyond
any experiment done so far.  Furthermore it seems unlikely that the
required performance can be demonstrated on Earth.  Therefore it is
desirable to consider a space experiment to demonstrate the required
level of performance.  

Several concepts have been proposed for a
mission to demonstrate technologies needed for LISA.  In 1999 NASA's
New Millennium Program supported a Phase A study of such a mission
concept.  The Disturbance Reduction System would have included two
test masses and a laser interferometer to measure the distance between
them.  If the forces on each test were small enough then the distance
between them would change very little, showing that the level of force
noise was near the LISA requirement.  The DRS concept was studied
along with two other mission concepts for demonstrating technologies
for future space science missions.  DRS was not selected for
implementation, primarily because the estimated cost the achieve the
DRS goals exceeded the cost cap.  (The selected mission was the
Nanosat Constellation Trailblazer: 
\htmladdnormallink{
http://nmp.jpl.nasa.gov/st5/}
{http://nmp.jpl.nasa.gov/st5/}).

Because of the high priority status of LISA with NASA and NRC reviews,
there are continuing efforts to find the best way to implement the
desired technology demonstration.  One possibility being very actively
studied is the possibility of adding a LISA Test Package to the ST3
Separated Spacecraft Interferometer mission
(\htmladdnormallink{
http://origins.jpl.nasa.gov/missions/st3.html}
{http://origins.jpl.nasa.gov/missions/st3.html}).  The ST3 mission is
planned for launch into an Earth-trailing orbit in 2005.  An
Earth-trailing orbit is very desirable for a LISA technology
demonstration because the thermal, magnetic, and gravitational
environment is much more stable than for an Earth-orbiting mission.
The environmental stability is key for achieving the low level of
forces needed for the LISA demonstration.  At this time, the
possibilities for allocating the required funding from NASA, and for
forming a partnership arrangement with ESA, for this option are being
actively pursued.

\vfill
\pagebreak

\section*{\centerline {
An update on the r-mode instability}}
\addcontentsline{toc}{subsubsection}{\it  
An update on the r-mode instability, by Nils Andersson}
\begin{center}
    Nils Andersson, University of Southampton, UK\\
\htmladdnormallink{N.Andersson@maths.soton.ac.uk}
{mailto:N.Andersson@maths.soton.ac.uk}
\end{center}

In the last two years the r-modes in rotating neutron stars
have attracted a lot of attention. The main reason for this is that they
are unstable due to the emission of gravitational waves via a mechanism 
that was discovered by Chandrasekhar, Friedman and Schutz 
more than 20 years ago.
Until recently the r-modes ---which are essentially horizontal currents
associated with very small density variations--- had  not been 
considered in this 
context. Hence the discovery that they are unstable at all rates 
of rotation in a perfect fluid star~[1]
came as a slight surprise. And even 
more of a surprise was the subsequent realization~[2]
that the unstable
r-modes (which radiate mainly through the current multipoles)
provide a much more severe constraint on the rotation rate of 
viscous stars (viscosity tends to counteract mode-growth due to
gravitational radiation) than the previously considered f-modes
(which are dominated by the mass multipoles). 
A direct comparison shows that
the f-mode becomes unstable when the star is spun up to roughly 95\% of the  
mass-shedding limit, while the dominant r-mode becomes unstable 
already at 5\% of the maximum spin rate (at some temperature).
With these early estimates, the r-mode instability emerged as a 
potential agent for spinning
nascent neutron stars down to rotation rates similar to the initial
period inferred for the  Crab pulsar ($P\approx 19$~ms),
in the process radiating an amount of gravitational
waves that should be detectable with LIGO II for sources in the Virgo
 cluster~[3].
It was also suggested~[4]
that the instability could operate in older, colder
neutron stars and perhaps explain the clustering 
of spin periods in the range 260-590~Hz of accreting neutron stars 
in Low-Mass X-ray Binaries (LMXB) indicated by the kHz QPOs.

Understandably, these possibilities created some excitement among workers
in this field and some 50 papers discussing the r-mode 
instability have since appeared. My intention here is to 
provide an update on the current status of this discussion without
going into too much detail. The interested reader is referred to an 
exhaustive review on the subject~[5]
(and the original papers, of course).
My main aim is to describe how our understanding of the r-modes and 
the associated instability has changed since the first studies. 
To some extent this is a simple task, because all the original ideas remain
relevant. No one has yet provided a demonstration that the mechanism cannot work
or that our original thinking was seriously flawed. This is, of course, 
good news. Less comforting is the fact that the questions that 
need to be addressed to make further progress are very hard and involve
a lot of essentially unknown physics. 

A natural point of departure for this survey is the case for r-modes
in hot young neutron stars
emerging from supernova explosions. A newly born neutron star
should cool to the temperature at which the dominant r-mode goes 
unstable (a few
times $10^{10}$~K) in a few seconds.
Provided that the star spins fast enough the r-mode will then 
grow with an ``e-unfolding'' time of a few tens of seconds until
it enters the nonlinear regime and... then what?  In the first studies
of the problem it was assumed that nonlinear effects (e.g. coupling to
other modes) would lead to the mode saturating at some 
large amplitude~[3]. The mode would continue to 
radiate away angular momentum
and the star would spin down from the mass-shedding limit to 
a period of 15-20~ms in a year or so~[2]. In these 
models a crucial 
parameter is the amplitude of saturation. In order for the instability
to have a dramatic effect on the spin-evolution of a young neutron star, 
the r-mode must be allowed to grow to a reasonably large amplitude. 
Intuitively, one might expect non-linear effects to become relevant
at much smaller mode-amplitudes than those considered in the early
work. However, the indications are now that the mode
will be able to grow surprisingly large. 
This is demonstrated by very recent 3D time-evolutions (using a 
fully nonlinear relativistic hydrodynamics code with 
the spacetime ``frozen'') of Stergioulas and Font~[6]. 
The first results of investigations into the nonlinear
coupling between r-modes and other modes 
seem to point in the same direction~[7]. 
There are no signs of mode-saturation until at very large amplitudes. 
It should, of course, be noted that much work remains to 
be done on this problem before we can draw any firm conclusions.

The original spin-down scenarios were based on the assumption that
the star  evolves along a sequence of uniformly
rotation equilibrium models as it  
loses angular momentum. Recent work indicates that this is unlikely to be 
the case.  One might 
expect that a large amplitude unstable mode will lead to differential
rotation in the stellar fluid.  It is well-known that
this is the case for the bar-mode instability
in the Maclaurin spheroids. Once spun up to the point where 
the bar-mode becomes unstable, the Maclaurin spheroids evolve
through a sequence of differentially rotating Riemann S-ellipsoids. 
One might expect an analogous evolution for stars governed by the 
r-mode instability. Evidence in favor of this possibility have been
presented by Rezzolla, Lamb and Shapiro~[8]. They argue that the r-mode
leads to a nonlinear differential drift of the various fluid elements. 
Their calculation is based on inferring higher order (in the mode-amplitude)
results from established linear results, and may not be quantitatively
reliable, but it provides an indication that nonlinear effects will 
severely alter the fluid motion. 
This result is supported both by the time-evolutions of Stergioulas
and Font~[6] and a shell toy model  
studied by Levin and Ushomirsky~[9]. In the latter case the nonlinear
effects can be determined exactly, and they lead to the anticipated
differential drift. Furthermore, the shell toy-model shows that, once
radiation reaction is implemented, another source of differential 
rotation comes into play. Thus it would seem almost certain that
differential rotation will play a key role in any realistic r-mode
scenario. Differential rotation immediately brings 
magnetic field effects into 
focus. While effects due to electromagnetic waves generated by an oscillation
mode are typically small~[10], 
differential rotation may lead to a twisting 
of the field lines and  a dramatic increase
in the field strength. In the case of the r-modes the instability
scenario may lead to the generation of a very strong toroidal magnetic
field~[8]. 

Following an original suggestion by Bildsten and Ushomirsky~[11], 
much work in the last eight months or so has been focussed
on the interface between the fluid core and the solid crust
in a slightly older neutron star. Given that the crust 
is  likely form already at a temperature of the order of $10^{10}$~K
this discussion is relevant for all but very young neutron stars.
Bildsten and Ushomirsky showed that a viscous boundary layer 
at the crust-core interface would lead to a very strong dissipation mechanism
that would prevent the instability from operating unless the 
rotation period was very short. The original estimates 
seemed to suggest that the r-mode instability would not be 
relevant in the LMXBs and that it would not be able to 
spin a newly born neutron star down to spin periods beyond
a few milliseconds. With more detailed studies these suggestions have 
been revised~[12],  and it now seems as if the instability could well
be relevant for the LMXBs (perhaps leading to a cyclical spin-evolution~[13]). But the uncertainties are large and many issues remain to 
be explored in this context.
The crust-core discussion has  led to suggestions that the heat 
released in the viscous boundary layer may, in fact,  melt
the crust. An interesting possibility, 
suggested by Lindblom, Owen and 
Ushomirsky~[14], is that the final outcome is a kind of mixed state, 
with ``chunks of crust'' immersed in the fluid. 
To estimate the mode-dissipation associated with such a  situation 
is, of course, very difficult. 
Also worth mentioning in this context are the results 
 of Wu, Matzner and Arras~[15].
They argue that
the crust-core
boundary layer is likely to be turbulent which would 
provide a mechanism for saturation. However, 
one can infer
that the resultant saturation amplitude is of order unity for 
rapidly rotating stars. This 
could well indicate that the modes saturate due to some alternative, 
as yet unspecified, mechanism. 

Progress on all these issues is somewhat hampered by the
lack of detailed quantitative results. It may be appropriate
to provide a contrast by concluding this 
discussion by emphasizing two particular cases where hard
calculations have provided  relevant results. 
The first of these concerns r-modes in superfluid
stars. This is an important issue since the bulk of a neutron star is expected
to become superfluid once it cools below a few times $10^9$~K.
At this point some rather exotic dissipation mechanisms come into play, and it
turns out (somewhat paradoxically) that a superfluid star is more 
dissipative than a normal fluid one. The most important new 
mechanism is the so-called mutual friction which has been shown 
to completely suppress the instability associated with the f-modes.
The initial expectations were that mutual friction would also have 
a strong effect on the r-modes~[2,3]. 
Detailed calculations by Lindblom 
and Mendell~[16] have shown that this is not necessarily the case. 
The outcome depends rather sensitively on the detailed superfluid model
(the parameters of the so-called entrainment effect), 
and only in a small set of the models considered by Lindblom and Mendell
do mutual friction affect the r-modes in a significant way. It would thus
seem as if the r-mode 
instability may prevail also in superfluid stars. 

Another important issue regards r-modes in fully relativistic stars. 
After all, the instability is a truly relativistic effect (being
driven by gravitational radiation) and a relativistic calculation
is required if we want to understand radiation reaction ``beyond 
the quadrupole formula''. And it should be recalled that
the quadrupole formula leads to a significant error (it deviates from the 
true result by 20-30\% already for $M/R\approx 0.03$) in 
estimates of the gravitational wave damping of the f-mode~[17]. 
Furthermore, it is known that relativistic effects
tend to further
destabilize the f-modes~[18]. While the quadrupole
f-mode does not become unstable below the mass-shedding 
limit in a Newtonian star it does so in the relativistic case.
For all these reasons
the modeling of relativistic r-modes is a crucial 
step towards improved estimates of the instability timescales.
It turns out that relativistic modes whose  dynamics
is mainly  determined by the Coriolis force generally
have a ``hybrid'' nature. This  makes the calculation rather complicated, but
significant progress on determining the relativistic analogue of the 
Newtonian r-modes has been made recently.  These results are detailed in 
 Lockitch's PhD thesis~[19], 
as well as a recent
paper~[20] where the
post-Newtonian corrections to the $l=m$ r-modes of uniform 
density stars are calculated. 
Estimates of the growth rate of the unstable modes in the 
fully relativistic case are currently being worked out as an
extension of this work.

At this point I hope it is clear that, despite some recent 
progress, the uncertainties regarding the astrophysical 
role of the r-mode instability remain considerable.
This is obviously somewhat disconcerting since it means
that our understanding of this mechanism is not detailed
enough to provide reliable theoretical templates
that can be used to search for the associated gravitational
waves in data taken by LIGO, GEO600, VIRGO or TAMA. 
In fact, I think it is quite
unlikely that theorists will be able to provide this kind of 
information any time soon.  
After all, a detailed understanding of the 
involved issues demands a successful modeling of a regime where many extremes
of physics meet.
In view of this, I believe the challenge is to invent a pragmatic detection 
strategy based on general principles rather than 
detailed theoretical information.  
After all, would it not be quite exciting if an actual detection would 
provide us with some of the missing pieces of this pulsar puzzle,
and help improve our understanding of  
general relativity, supranuclear physics, magnetic fields, superfluidity
etcetera?

\def\prl#1#2#3{{ Phys. Rev. Lett.\ }, {\bf #1}, #2 (#3)}
\def\prd#1#2#3{{ Phys. Rev. D}, {\bf #1}, #2 (#3)}
\def\plb#1#2#3{{ Phys. Lett. B}, {\bf #1}, #2 (#3)}
\def\prep#1#2#3{{ Phys. Reports}, {\bf #1}, #2 (#3)}
\def\phys#1#2#3{{ Physica}, {\bf #1}, #2 (#3)}
\def\jcp#1#2#3{{ J. Comput. Phys.}, {\bf #1}, #2 (#3)}
\def\jmp#1#2#3{{ J. Math. Phys.}, {\bf #1}, #2 (#3)}
\def\cpr#1#2#3{{ Computer Phys. Rept.}, {\bf #1}, #2 (#3)}
\def\cqg#1#2#3{{ Class. Quantum Grav.}, {\bf #1}, #2 (#3)}
\def\cma#1#2#3{{ Computers Math. Applic.}, {\bf #1}, #2 (#3)}
\def\mc#1#2#3{{ Math. Compt.}, {\bf #1}, #2 (#3)}
\def\apj#1#2#3{{ Astrophys. J.}, {\bf #1}, #2 (#3)}
\def\apjl#1#2#3{{ Astrophys. J. Lett.}, {\bf #1}, #2 (#3)}
\def\apjs#1#2#3{{ Astrophys. J. Suppl.}, {\bf #1}, #2 (#3)}
\def\acta#1#2#3{{ Acta Astronomica}, {\bf #1}, #2 (#3)}
\def\sa#1#2#3{{ Sov. Astro.}, {\bf #1}, #2 (#3)}
\def\sia#1#2#3{{ SIAM J. Sci. Statist. Comput.}, {\bf #1}, #2 (#3)}
\def\aa#1#2#3{{ Astron. Astrophys.}, {\bf #1}, #2 (#3)}
\def\apss#1#2#3{{Astrop. Sp. Sci.}, {\bf #1}, #2 (#3)}
\def\mnras#1#2#3{{ Mon. Not. R. Astr. Soc.}, {\bf #1}, #2 (#3)}
\def\prsla#1#2#3{{ Proc. R. Soc. London, Ser. A}, {\bf #1}, #2 (#3)}
\def\ijmpc#1#2#3{{ I.J.M.P.} C {\bf #1}, #2 (#3)}

\par
{\em{References:}}
\par
[1]~~N. Andersson \apj{502}{708}{1998};
	J.L. Friedman and  S.M. Morsink \apj{502}{714}{1998}
\par
[2]~~ L. Lindblom, B.J. Owen and S.M. Morsink
	\prl{80}{4843}{1998}; N.W Andersson, K.D. Kokkotas and
B.F. Schutz \apj{510}{846}{1999}
\par
[3]~~B.J. Owen, L. Lindblom, C. Cutler, B.F. Schutz, A. Vecchio and
	N. Andersson \prd{58}{084020}{1998}
\par
[4]~~L. Bildsten \apjl{501}{89}{1998}; 
	N. Andersson, K.D. Kokkotas and N. Stergioulas \apj{516}{307}{1999}
\par
[5]~~N. Andersson and K.D. Kokkotas
to appear in Int. J. Mod. Phys. D (2000)
\par
[6]~~N. Stergioulas and J.A. Font {\em Nonlinear r-modes in 
rapidly rotating relativistic stars} preprint astro-ph/0007086
\par
[7]~~S. Morsink private communication
\par
[8]~~L. Rezzolla, F.K. Lamb and S.L. Shapiro \apjl{531}{139}{2000}
\par
[9]~~Y. Levin and G. Ushomirsky
{\em Nonlinear r-modes in a spherical shell: issues of principle}
preprint astro-ph/9911295
\par
[10]~~W.C.G. Ho and D. Lai {\em R-mode oscillations and spindown of 
young rotating magnetic neutron stars} preprint astro-ph/9912296
\par
[11]~~L. Bildsten  and G. Ushomirsky \apjl{529}{33}{2000}
\par
[12]~~N. Andersson, D.I. Jones, K.D. Kokkotas  and N. Stergioulas
	\apjl{534}{75}{2000}; S. Yoshida and U. Lee {\em
R-modes of neutron stars with a solid crust} preprint astro-ph/0006107; 
Y. Levin and G. Ushomirsky
{\em Crust-core coupling and r-mode damping in neutron stars: a toy model}
preprint astro-ph/0006028
\par 
[13]~~Y. Levin \apj{517}{328}{1999}
\par
[14]~~ L. Lindblom, B.J. Owen and G. Ushomirsky
{\em Effect of a neutron-star crust on the r-mode instability}
preprint astro-ph/0006242
\par
[15]~~Y. Wu, C.D. Matzner and P. Arras
{\em R-modes in Neutron Stars with Crusts: Turbulent Saturation, Spin-down, and Crust Melting} preprint astro-ph/0006123
\par
[16]~~L. Lindblom and  G. Mendell \prd{61}{104003}{2000}
\par
[17]~~E. Balbinski, B.F. Schutz, S. Detweiler and  L. Lindblom 
\mnras{213}{553}{1985}
\par
[18]~~N. Stergioulas, Living Reviews in Relativity, 
	1998-8 (1998) \\ 
\htmladdnormallink
{http://www.livingreviews.org/Articles/Volume1/1998-8stergio/}
{http://www.livingreviews.org/Articles/Volume1/1998-8stergio/}
\par
[19]~~K.H Lockitch  {\em 
Stability and rotational mixing of modes in Newtonian and relativistic stars}
Ph.D. Thesis, University of Wisconsin - 
	Milwaukee (1999). Available as preprint gr-qc/9909029
\par
[20]~~K.L. Lockitch, N. Andersson and J.L. Friedman
to appear in  Phys. Rev. D  (2000)

\vfill
\pagebreak

\section*{\centerline {
Laboratory Experiments: A 14 ppm G measurement,}\\
\centerline{ a new sub-mm gravity
constraint,}\\\centerline{and other news from  MG9}}
\addcontentsline{toc}{subsubsection}{\it Laboratory experiments: news from MG9,
by Riley Newman}
\begin{center}
    Riley Newman, University of California, Irvine\\
\htmladdnormallink{rdnewman@uci.edu}
{mailto:rdnewman@uci.edu}
\end{center}

{\bf G to 14 ppm.}  An elegant new G measurement by Jens Gundlach and
Stephen Merkowitz of the U. Washington ``E\"{o}t-Wash" group was
reported at the April APS meeting in Long Beach, and at the Marcel
Grossmann meeting (MG9) in Rome this summer.  The reported result, $G=
(6.674215 \pm 0.000092)\times 10^{-11} m^3 kg^{-1} s^{-2}$, carries an
assigned uncertainty two orders of magnitude smaller than the 1500 ppm
uncertainty associated with the current recommended ``CODATA" G value
(the CODATA uncertainty reflects large discrepancies in G values
reported in the last decade -- see MOG Number 13).  Gundlach's
measurement has a number of novel features.  A PRL paper in press and
available in preprint form [1] describes the experiment.  At the heart
of the apparatus is a torsion balance placed on a turntable located
between a set of attractor spheres.  The turntable is first rotated at
a constant rate so that the pendulum experiences a sinusoidal torque
due to the gravitational interaction with the attractor masses.  A
feedback is then turned on that changes the rotation rate so as to
minimize the torsion fiber twist.  The resulting angular acceleration
of the turntable, which is now equal to the gravitational angular
acceleration of the pendulum, is determined from the second
time-derivative of the turntable angle readout.  Since the torsion
fiber does not experience any appreciable deflection, this technique
is independent of many torsion fiber properties, including
anelasticity, which may have led to a bias in previous measurements.
The attractor masses revolve around the pendulum on a second turntable
whose constant angular velocity differs from that of the pendulum's
turntable.  This motion of the attractor masses makes their torque on
the pendulum readily distinguishable from torque due to ambient
lab-fixed gravitational fields.  Another key feature described in the
forthcoming paper and earlier papers [2,3] is a pendulum in the form
of a thin rectangular plate.  The gravitational torque on the pendulum
is dominantly determined by the ratio of its quadrupole moment to
moment of inertia -- a ratio which is independent of the shape and
mass distribution of the pendulum in the limit that it has negligible
width.  This greatly eases the metrology requirement for the pendulum,
in contrast to earlier experiments where pendulum metrology has been a
limiting factor.

{\bf G at MG9.} A session at MG9 was devoted primarily
to G measurements, several of which target accuracy comparable to that
of Gundlach and Merkowitz.  Gundlach reported the measurement
described above.  Tim Armstrong reported measurements made at the New
Zealand Measurement Standards Laboratory using a torsion pendulum in
two modes: one in which a servo system and rotating platform ensured
that there was no significant fiber twist, yielding $G= (6.6742 \pm
0.0007)\times 10^{-11} m^3 kg^{-1} s^{-2}$ [4], and a more recent one
using the dynamic (``time-of-swing") method yielding $G= (6.675 \pm
0.01)\times 10^{-11} m^3 kg^{-1} s^{-2}$.  The latter value has much
larger uncertainty but is consistent with the former, and both values
are consistent with that of Gundlach and Merkowitz.  Jun Luo described
a new G measurement being developed by his lab in China, which should
improve on his measurement published recently [5]: $G= (6.6699 \pm
0.0007)\times 10^{-11} m^3 kg^{-1} s^{-2}$.  Stephan Schlamminger gave
a progress report on the University of Z\"{u}rich G measurement using
a beam balance and mercury-filled steel tank source masses.  This
experiment [6], which has been troubled in the past by systematic
error, shows encouraging progress toward a 10 ppm measurement.  Jim
Faller reported progress of a G determination based on measurement of
the differential deflection of a pair of suspended masses which form a
Fabry-Perot cavity; this experiment expects 50 ppm G accuracy,
significantly improving on an earlier G measurement by Faller's group
[7].  Michael Bantel reported progress of the UC Irvine G measurement
using a high-Q cryogenic torsion pendulum operating in the dynamic
(``time-of-swing") mode.  Ho Jung Paik described his proposed
cryogenic G measurement in which a set of four magnetically suspended
test masses would be located symmetrically on the periphery of a
slowly rotating turntable. Paik's determination of G would be made by
measuring the turntable rotation speed required to keep the masses at
a fixed radius when an attracting mass is lowered into the center of
the array of test masses.  Paik's proposed experiment targets 1 ppm
accuracy.  In the one non-G talk of the session, Andrej \v{C}ade\v{z}
with Jurij Kotar described the University of Ljubljana inverse square
law test, in which two pairs of source masses rotate continuously
about a torsion pendulum -- one pair at opposite 971 mm radii and
another at 383 mm radii along an axis perpendicular to that of the
first pair.  The masses of the pairs are chosen to produce null
pendulum excitation at twice the rotation frequency for a Newtonian
force law.  The group expects to improve on their previous limit [8]
which constrained a Yukawa interaction term to be $(-0.2 \pm 6) \times
10^{-3}$ relative to gravity over a distance range 0.2 m to 0.45 m.
     
It seems increasingly clear that the anomalous PTB G measurement [9]
must be in error.  However, new measurements have yet to converge
satisfactorily.  At the ``CPEM2000" metrology conference in Australia
in May this year, a BIPM group led by Terry Quinn reported
(\underline{preliminary}) results of G measurements using a torsion
pendulum suspended by a strip fiber.  Such a pendulum is minimally
subject to systematic error associated with fiber anelasticity,
because the dominant part of its effective torsion constant is
gravitational in origin and hence lossless.  The measurements were
made in two modes: 
an unconstrained static measurement,
yielding $G= (6.6693 \pm 0.0009)\times
10^{-11} m^3 kg^{-1} s^{-2}$ and a static measurement in which the
pendulum was servoed to zero displacement with a calibrated
electrostatic force, yielding $G= (6.6689 \pm 0.0014)\times 10^{-11}
m^3 kg^{-1} s^{-2}$.  The two methods yield consistent results which
are however more than 5 of their own standard deviations from the G
value obtained by Gundlach and Merkowitz.

{\bf Sub-mm gravity at MG9.}  The highlight of the MG9 session on
short-range tests of gravity was preliminary results of
the ``E\"{o}t-Wash" group's test, reported by Jens Gundlach.  The
instrument of this experiment is a torsion pendulum in the form of a
horizontal disk with ten holes arranged symmetrically azimuthally,
suspended above a rotating attractor, with a thin copper electrostatic
shield between.  The attractor is in the form of two copper disks,
each with a set of ten holes.  The lower of these rotating disks has a
fixed angular displacement relative to the upper and is more massive,
arranged so that for a particular pendulum-attractor separation the
pendulum experiences no torque modulation at the signal frequency of
ten times the attractor rotation frequency if gravity is Newtonian.
Gundlach presented preliminary results in the form of a sketched plot
indicating a one sigma limit on the order of 2\% of gravity at a range
of about 1.5 mm.  The test is expected to yield still better
constraints soon.

John Price reported a current sensitivity about 100 times
gravitational strength at 0.1 mm, expected to improve to 1 times
gravitational strength at that distance using his existing room
temperature instrument and to improve still further with a planned
cryogenic instrument.

Michael Moore discussed the short-range test he is developing with
Paul Boynton, which uses a near-planar torsion pendulum suspended
above a near-planar source mass, configured to give a nearly null
signal for purely Newtonian gravity.  The expected sensitivity of
their apparatus to an anomalous force is about 0.25 of gravity at 0.25
mm and 0.01 of gravity at 1 mm.

Aharon Kapitulnik described his present cantilever-based instrument at
Stanford, which has projected sensitivity better than .05 of gravity
at 0.08 mm, and discussed possible future improvements.

Giuseppe Ruoso discussed the apparatus of the Padua group.  Currently
optimized for Casimir force measurements, the instrument does not yet
have good sensitivity for short-range gravity measurements.  When the
Casimir tests are completed the group expects to optimize it for
gravity tests, and expects sensitivity on the order of $\alpha = 10^7$
to $10^8$ for ranges of a few microns, in a yet-unexplored region of
the $\alpha - \lambda $ plane.

Ho Jung Paik reported the design of a cryogenic null test of the
inverse-square law, with expected sensitivity at a level 0.0001 of
gravity at 1 mm and 0.01 of gravity at 0.1 mm.

Ephraim Fischbach reviewed the motivations for short-distance gravity
tests, and discussed prospects for very short range tests using atomic
force microscopy.  Dennis Krause as well as Ephraim discussed ways of
dealing with the severe problems of molecular background forces in
extremely short range tests.

Christian Trenkel reported the development of a torsion balance using
a Meissner effect suspension, and this instrument's prospective
applications in weak force physics such as a spin-mass coupling
experiment.

A list including other current mm-scale gravity tests, with a little
more detail on some of the projects reported above, is available in
MOG number 15.

{\bf Laboratory equivalence principle tests at MG9.}  A session on
equivalence principle tests, chaired by Ramanath Cowsik, included
talks on both space and laboratory tests; I report here only on the
latter.

Nadathur Krishnan reported the status of the TIFR equivalence
principle experiment, which employs a torsion pendulum with a 3.6
meter long torsion fiber of rectangular cross section, operating in a
chamber deep underground in a seismically very quiet region of India.
The test operates in a Dicke mode, using the sun as acceleration
source, targeting a sensitivity at a level of $\eta \approx 10^{-13}$.
Continuous operation of the instrument is about to begin.

Paul Boynton discussed a novel mode in which a torsion pendulum may be
used to measure anomalous forces, based on measurement of the second
harmonic component of the pendulum's oscillatory motion.  This method,
introduced by Michael Moore in Paul's group, has the great advantage
that it is extremely insensitive to variation of the fiber
temperature, in contrast to force measurements based on measurement of
a pendulum's oscillation frequency or static displacement.

I gave a short talk on prospects for improved terrestrial equivalence
principle tests using a cryogenic torsion pendulum, taking advantage
of the high Q and good temperature control achievable with such an
instrument.  In principle such an instrument should be capable of
$\eta$ sensitivities of $10^{-14}$ or better, although many practical
difficulties are to be encountered.

Wolfgang Vodel gave a progress report on the Bremen Drop Tower test of
the equivalence principle, in which a superconducting differential
accelerometer falls 109 meters in an evacuated tube.  This system is
expected to be capable of $\eta$ sensitivity at a $10^{-14}$ level,
with a theoretical limit at a $10^{-16}$ level and a near-term result
anticipated at a $10^{-13}$ level.

Cliff Will reviewed tests of the three ingredients of the Einstein
Equivalence Principle -- universality of free fall, local Lorentz
invariance, and local position invariance -- and discussed their
theoretical implications.

{\bf References:}

[1] Jens Gundlach and Stephen Merkowitz, PRL in press, preprint at
\htmladdnormallink{http://xxx.lanl.gov/format/gr-qc/0006043}
{http://xxx.lanl.gov/format/gr-qc/0006043}.

[2] J.H. Gundlach, E.G. Adelberger, B.R. Heckel and H.E. Swanson, ,
Phys.  Rev.  D{\bf 54}, R1256 (1996)

[3] J.H. Gundlach, Meas. Sci. Technol. {\bf 10}, 454 (1999)

[4] M.P. Fitzgerald and T.R. Armstrong, Meas.  Sci.  Technol. {\bf
10}, 439 (1999)

[5] Jun Luo et al., Phys.  Rev.  D{\bf 59}, 042001 (1998)

[6] F. Nolting, J. Schurr, S. Schlamminger and W. K\"{u}ndig,
Meas. Sci. Technol. {\bf 10}, 487 (1999)

[7] J.P. Schwarz, D.S. Robertson, T.M. Niebauer and J.E. Faller,
Meas. Sci. Technol. {\bf 10}, 478 (1999)

[8] A. Arn\v{s}ek and A. \v{C}ade\v{z}, Proceedings of the 8th Marcel
Grossmann Meeting, 1174 (World Scientific, 1999)

[9] W. Michaelis, H. Haars, and R. Augustin, Metrologia {\bf 32}, 267 (1996)

\vfill
\pagebreak

\section*{\centerline {Progress toward Commissioning the LIGO detectors}}
\addcontentsline{toc}{subsubsection}{\it 
Progress toward Commissioning the LIGO detectors, by Stan Whitcomb}
\begin{center}
    Stan Whitcomb, LIGO Laboratory\\
\htmladdnormallink{stan@ligo.caltech.edu}
{mailto:stan@ligo.caltech.edu}
\end{center}

This past year has seen great progress in two distinct detector activities:
installation--defined as getting working subsystems into position--and
commissioning--making the subsystems work together to operate as a
complete detector with full sensitivity.

Our installation and commissioning plan has evolved into one where
each of the three interferometers has a well-defined role, and the
scheduling of the work on each one has been tailored to its role.
The Hanford 2 km interferometer is the first in line and serves as
a "pathfinder" to identify problems early. The Livingston 4 km
interferometer follows about 6 months behind, and is used for
problem resolution and detailed characterization. We will initiate
coincidence testing as soon as the first two interferometers are
operational, but we will deliberately delay installation of some
elements of the Hanford 4km interferometer (primarily control
electronics) to enable lessons learned from the first two interferometers
to be realized in redesign before installation. The LIGO I science
run will begin when reliable and calibrated coincidence data on
three interferometers can be taken while keeping the configuration
stable for substantial periods of time. The improvements to reach
final design goals in sensitivity and reliability will be alternated with
data running in a way that optimizes both the early running and
obtains integrated high sensitivity data taking before the completion
of the initial LIGO science run.

On the installation front, the fabrication of the detectors was completed
(with the exception of some electronics components), and most detector
components have been delivered to the Observatories for installation.
Installation of the Hanford 2 km interferometer was completed in May 2000.
Installation of the Livingston 4 km interferometer is being completed as this
Newsletter goes to press (September 2000).  As mentioned above the
Hanford 4 km interferometer installation has been intentionally delayed,
but substantial progress has been made: the seismic isolation has been
installed and the infrastructure (networking, data acquisition, monitoring
equipment) has been installed and tested.

Commissioning the LIGO detectors began even before the installation was
complete.  On both the Hanford 2 km interferometer and the Livingston 4 km
interferometer, the pre-stabilized laser has been integrated with the
mode cleaner (a suspended cavity to stabilize the laser beam before it
enters the interferometer). Initial characterization of the laser/mode cleaner
system has been completed and show that the combination is already
very close to meeting their performance requirements.

In December 1999, we began a four month test of the 2 km interferometer
in which each arm of the interferometer was separately locked to the laser.
This test was performed to measure optical properties of the arms, to test
the interferometer sensing and control electronics, to gain information about
the environmental noise sources and to exercise the data acquisition and
control networks. Lock sections up to 10 hours were obtained and all planned
investigations were successfully concluded.  At the end of the testing, a 
24-hour
stretch of data was taken and archived for use by groups developing
software and techniques for data analysis and detector characterization.

During the past summer, we have gradually begun to bring the entire 2 km
interferometer on-line.  We began by operating it in the recycled Michelson
configuration, (without the long arm cavities).  This simple configuration 
allowed us
to test the control systems: verifying the myriad connections, measuring
transfer functions, setting modulation/demodulation phases, all of the nuts 
and bolts
of precision interferometry that must be in place before everything will work.
Most recently, we have locked the power-recycled Michelson with one
Fabry-Perot arm also locked on resonance.  This initiated another set of
control system measurements which should lead to locking the full 
interferometer
early this fall.

Of course, a number of problems have been encountered along the way, but many
have been solved, solutions to others are in the works, and none will 
jeopardize
the performance or schedule significantly.  We are on track to initiate the
first triple coincidence science runs by early 2002.

\vfill
\pagebreak

\section*{\centerline {
160 Hours of Data Taken on the TAMA300}\\
\centerline{ Gravitational Wave Detector}}
\addcontentsline{toc}{subsubsection}{\it 
160 Hours of Data Taken on TAMA300, by Seiji Kawamura}
\begin{center}
    Seiji Kawamura, National Astronomical Observatory of Japan\\
\htmladdnormallink{seiji.kawamura@nao.ac.jp}
{mailto:seiji.kawamura@nao.ac.jp}
\end{center}

The TAMA project, the Japanese effort for detecting gravitational waves
using the 300m laser interferometer, successfully took 160 hours of data
between August 21, 2000 and September 4, 2000. The best sensitivity of the
detector was about $5\times10^{-21} {\rm Hz}^{-1/2}$ around 1 kHz in 
terms of strain, which gives a signal-to-noise ratio of 
20 to 30 for gravitational waves emitted
from a binary neutron star coalescence in the center of our galaxy. The
interferometer was operated remarkably stably; the longest continuous
locking time was more than 12 hours, and on one day it was locked for more
than 23 hours out of 24 hours. The quality of the data was also drastically
improved compared with our previous runs. First the non-stationary noise
which appeared very often in the previous data runs was significantly
reduced. Secondly approximately 100 signals including feedback and error
signals of various control loops and environmental signals such as ground
motion were also recorded so that any spurious signals in the interferometer
output can be vetoed by correlating them with other channels. The obtained
data are now being analyzed for gravitational wave detection as well as for
diagnosis purposes of the interferometer. We will further improve the
sensitivity and stability of the detector from now until in January 2001 we
plan to hold a two-month data run.

Please have a look at our home page.
\htmladdnormallink{http://tamago.mtk.nao.ac.jp/}
{http://tamago.mtk.nao.ac.jp/}

Also the "Data treatment guideline of TAMA" 
can be found at the following web
site.
\htmladdnormallink{
http://tamago.mtk.nao.ac.jp/tama/data-access.html}
{http://tamago.mtk.nao.ac.jp/tama/data-access.html}

\vfill\eject

\section*{\centerline {
Kipfest}}
\addtocontents{toc}{\protect\medskip}
\addtocontents{toc}{\bf Conference reports:}
\addtocontents{toc}{\protect\medskip}
\addcontentsline{toc}{subsubsection}{\it  
Kipfest, by Richard Price}
\begin{center}
    Richard Price, University of Utah\\
\htmladdnormallink{rprice@mail.physics.utah.edu}
{mailto:rprice@mail.physics.utah.edu}
\end{center}

With the development of time machines apparently bogged down,
Caltech's Kip Thorne had little alternative but to reach his 60th
birthday on June 1, 2000. A large corps of relativity observers were
on hand for the occasion to mark the moment, and to prevent
Kip from relaxing and enjoying it.  A set of Kip's former students
organized a three day "KipFest" on June 1 --3 that included two days
of scientific talks, a day of popular talks for the public, and a
banquet.

The scientific talks, presented on June 1 and June 2, were not
primarily conference-style reports on scientific news, but rather were
talks emphasizing the scope of Kip Thorne's contributions to various
branches of relativity and relativistic astrophysics. They included
overviews of problems on which Kip had worked and reminiscences
about Kip.  The full two-day scientific program can be found
linked to the KipFest website at

\htmladdnormallink{
  http://wugrav.wustl.edu/People/CLIFF/KipFest/kipmain.html }
{  http://wugrav.wustl.edu/People/CLIFF/KipFest/kipmain.html }

The list of topics is striking in its breadth of topics. Kip had been
a driving force, or major contributor, to problems ranging from
experimental approaches to gravity (e.g., the talks by Rai Weiss and
Vladimir Braginsky) to wormholes (Eanna Flanagan's talk) and "real"
astrophysics (talks by Roger Blandford and Anna Zytkow). Even with all
the memorable moments of the two days, one moment stands out. Jim
Hartle related how he had worked with Kip on slowly rotating stars
more than 20 years ago, but then was lured away by the siren call of
quantum gravity. (Jim suggested that he could be accused of
"not having done a lick of honest work since.") The call interrupted
work on a final Hartle-Thorne paper that Kip had started and Jim was
to finish. The paper had remained unfinished, in the back of Jim's
filing cabinet. But only until June 2, when Jim ended his talk by
handing Kip the final draft!

To honor Kip Thorne's commitment to bringing exotic physics to
non-scientists, five talks were presented on Saturday, June 3, by speakers
with a gift for communicating the ideas of science. These talks, free
to the public, were held in Caltech's Beckman auditorium, and
attracted over a thousand listeners.  Stephen Hawking and Igor Novikov
discussed wormholes and time travel, and Kip Thorne made predictions
for what lay ahead in our field in the coming decade or so.  There
were also two talks not directly dealing with specific scientific
questions. The well known science writer Timothy Ferris talked about
the problem of communicating science to the public, and Alan Lightman,
who is both a scientist and a novelist, gave his insights about the
different kinds of creativity involved in his two careers.

It is not really possible to describe the banquet on Friday evening.
You had to be there. There were a few of the short speeches that one
expects, most notably by John Wheeler. But there was a somewhat
unexpected reminiscence by football/TV star Merlin Olsen, about the
early scientific curiosity of his boyhood friend Kip. (It had to do
with frogs, non-relativistic frogs.) Kip's sisters shared other
memories of his early days, and Kip did a wonderful job of hiding his
discomfiture.  Tradition grew yet thinner as Linda Williams, the
"Physics Chanteuse," celebrated physics and Kip with music. Undeterred
by the high musical standards she had set, the physics singing group
"Bernie and the Gravitones" went to the stage to make a rare
appearance and to make fools of themselves, an endeavor in which they
were judged to have been completely successful. The group was four of
Kip's former graduate students (Sandor Kovacs, Richard Price, Bernie
Schutz and Cliff Will) singing, in "the average key of B and a quarter
flat," about their "Wise Old Advisor from Pasadena" to a Jan and Dean
song from 1964.  The big finish of the evening was a presentation to
Kip of his academic family tree showing how he had populated the field
with 42 academic children (PhD's with Kip), how they in turn had
produced 70 academic grandchildren, and how they had produced 48
academic great grandchildren.

May the numbers continue to grow.

\vfill\eject

\section*{\centerline {Third Capra Meeting on Radiation Reaction}}
\addcontentsline{toc}{subsubsection}{\it  
Third Capra meeting, by Eric Poisson}
\begin{center}
    Eric Poisson, University of Guelph\\
\htmladdnormallink{
poisson@physics.uoguelph.ca}
{mailto:poisson@physics.uoguelph.ca}
\end{center}

The Capra series of meetings on radiation reaction in curved spacetime
were initiated in 1998 by Patrick Brady. The first meeting was held at
Frank Capra's ranch in southern California, and the name stayed, even
though the location of the meeting has changed. (Frank Capra is the
famous movie director who made such films as ``It's a wonderful life''
and ``Mr.~Deeds goes to town''. Capra had studied at Caltech before
going to Hollywood; he bequeathed his ranch to his alma mater, which
passed to Caltech upon his death.) The second meeting was held in
Dublin, Ireland, and was organized by Adrian Ottewill. This latest
installment was held at Caltech June 5--9, 2000, and was organized by
Lior Burko and Scott Hughes. Here I will present a rather broad
overview of the main issues discussed during the meeting, and
highlight just a few of the contributions. The complete proceedings
--- a copy of the slides presented by all the speakers --- can be
found at the meetings's web site: \htmladdnormallink{
http://www.tapir.caltech.edu/\~{}capra3/}
{http://www.tapir.caltech.edu/\~{}capra3/}.

This series of meetings is concerned with the motion of a small mass
in a strong gravitational field. It is known that in the limit of
vanishing mass, the particle moves on a geodesic of the background
spacetime. Away from this limit, however, the motion in the background
is no longer geodesic, and can be described in terms of a
self-force. (In some sense, the motion is geodesic in the perturbed
spacetime, which consists of the background plus the perturbation
created by the particle. For a point particle, however, the
perturbation is singular at the particle's location, and careful
thought must be given to the removal of the singular part of the field,
which does not affect the motion. In flat spacetime, this subtraction
gives rise to the well-known half-retarded minus half-advanced
potential.) The main focus of the meeting was the practical
computation of this force. 

While this problem raises many interesting issues of principle (such
as the removal of the singular part of the metric perturbation created
by a point particle), there is also a practical necessity. The
detailed modeling of gravitational waves produced by a solar-mass
compact object orbiting a massive black hole requires an accurate 
representation of the orbital motion, which evolves as a result of
radiation loss. In the generic case involving a rapidly rotating black
hole, this evolution must be calculated on the basis of a
radiation-reaction force. Such sources of gravitational waves will be
relevant for the Laser Interferometer Space Antenna (LISA), a
space-borne detector designed to measure low-frequency waves (it has a
peak sensitivity at around 1 mHz).       

The electromagnetic analogue to this problem was solved in 1960 by
DeWitt and Brehme [1], who derived a curved-spacetime expression
for the self-force acting on a point electric charge. The gravitational
self-force was obtained much more recently, first by Mino, Tanaka, and
Sasaki [2], and then by Quinn and Wald [3]. There is also a
similar force in the case of scalar radiation, which was calculated by
Quinn [4]. In all three cases the self-force is expressed as an
integral over the past world-line of the particle, and the integral
involves the nonsingular part of the retarded Green's function, which
has support inside the past light cone of the particle's current
position. The explicit evaluation of only this part of the Green's
function is challenging, however, and a good portion of the meeting
was devoted to this issue.            

A plausible method for calculating the Green's function involves a
separation-of-variable approach made possible by the symmetries of the
black-hole spacetime. (Thus far, all calculations have been restricted
to the case of a Schwarzschild black hole). It is a simple matter to
derive and solve the ordinary differential equation that governs each
mode of the Green's function. The problem lies with the fact that the
sum over all modes doesn't converge. (This is essentially because the
individual modes do not distinguish between the singular and
nonsingular parts of the Green's function.) Amos Ori, Leor Barack, and
Lio Burko [5] have devised a way of regulating the mode sum, so
as to extract something meaningful. Their results for simple
situations involving scalar radiation were presented at the meeting,
and are extremely promising. A similar regularization method was used
by Carlos Lousto [6], who calculated the gravitational self-force
acting on a radially infalling particle in Schwarzschild spacetime.  
Regularization was also exploited by Hiroyuki Nakano and Yasushi Mino
to calculate the gravitational self-force in the weak-field limit.    

Insight into the self-force problem can be gained by adopting a more
local point of view, and focusing on the immediate vicinity of the
particle. Such an approach permits a clear identification of the
singular part of the particle's field, which can then be decomposed
into modes and subtracted from the full field. Such a strategy was
adopted by Steve Detweiler (in the gravitational case) and by Patrick
Brady (in the scalar case). A variation on this theme is to start with
the Mino {\it et al.}~expression for the gravitational self-force
[2], and to evaluate the contribution to the world-line integral
that comes from the particle's very recent past. Results along those
lines were presented by Warren Anderson.     

The third Capra meeting has shown that the radiation-reaction problem
is progressing very nicely. There are still many issues left to sort
out, but it is nice to see that concrete results have now been
obtained. I expect that progress will be swift in the coming year, and
that the fourth (perhaps last?) meeting will be just as exciting as
the preceding ones.

[1] B.S. DeWitt and R.W. Brehme, Ann. Phys. (NY) {\bf 9}, 220
(1960). 

[2] Y. Mino, M. Sasaki, and T. Tanaka, Phys. Rev. D {\bf 55},
3457 (1997). 

[3] T.C. Quinn and R.M. Wald, Phys. Rev. D {\bf 56}, 3381
(1997). 

[4] T.C. Quinn, gr-qc/0005030. 

[5] See, for example, L.M. Burko, Phys. Rev. Lett. {\bf 84},
4529 (2000), and L. Barack, gr-qc/0005042. 

[6] C.O. Lousto, Phys. Rev. Lett. {\bf 84}, 5251 (2000). 

\vfill\eject

\section*{\centerline {
General Relativity and Quantum Gravity, at the } \\
\centerline{XIIIIth International Congress on Mathematical Physics}}
\addcontentsline{toc}{subsubsection}{\it  
GR at the XIIIth Congress on Mathematical Physics, by Abhay Ashtekar}
\begin{center}
    Abhay Ashtekar, Penn State\\
\htmladdnormallink{
ashtekar@gravity.phys.psu.edu}
{mailto:ashtekar@gravity.phys.psu.edu}
\end{center}

The International Association of Mathematical Physics hosts a
tri-annual congress to review the recent developments in the
field. The 13th congress in this series took place at Imperial
College, London, from 17th to 22nd July 2000. While quantum field
theory and statistical mechanics have been the major components of
these conferences, general relativity and quantum gravity have been
well represented at least since the early eighties and, over the
years, interest in sessions in our field has steadily increased.

At the London conference, Gerhard Huisken gave a plenary lecture on
{\it Energy inequalities for isolated gravitating systems} in which he
presented the recent proofs of the Penrose inequality (in the case of
a maximal slice).  Roughly, the inequality says that the total mass
should be greater than the square-root of the area of the apparent
horizon and thus strengthens the positive mass theorems proved in the
late seventies.  It was a lucid presentation of deep results, much
appreciated also by participants outside general relativity.  In
addition, there were two invited sessions.  The first talk in the
classical gravity session was given by Lars Andersson in which he
summarized recent results on approach to singularities in general
relativity coupled to a scalar field.  In a well-defined sense, the
scenario put forward by Belinskii, Khalatnikov and Lifshitz (BKL) in
the early sixties can now be rigorously justified in this case.  In
the second talk, Piotr Bizon first gave a succinct and exceptionally
clear review of the ``critical phenomena'' first discovered by
Choptuik and then summarized recent work which shows that many of
the key features arise already in simpler dynamical systems and are
thus not unique to Einstein's equations.

In the invited session on quantum gravity, John Barrett provided an
overview of the state sum models, emphasizing the use of combinatorial
methods and bringing out relation between diverse ideas that have come
from mathematics and physics. John Baez summarized the recent results
on black hole entropy based on the quantum geometry of isolated,
non-rotating horizons. Although the subject involves rather technical
ideas from diverse fields, he demonstrated his exceptional skill at
zeroing-in on the essentials and making everything fit together
naturally. In addition, there were two contributed sessions which were
also well attended. The classical gravity session emphasized recent
mathematical results on black holes. In the quantum gravity session,
while the first two talks were on ``standard'' mathematical physics
topics on the interface of general relativity and quantum physics, the
last two were on the interface between quantum gravity, philosophy of
science and quantum computing. Unfortunately, this attempt to broaden
and reach out to neighboring field did not succeed; there was a marked
difference in the level of precision and emphasis between the two sets
of talks. Finally, there was a poster session which contained a number
of exceptionally interesting presentations.

In addition to these sessions which Peter Aichelburg and I organized,
there were other activities related to gravitational physics. In
particular, there were two round-table discussions. The first was on
{\it Quantum theory of space-time}, organized by Chris Isham and
chaired by John Klauder, in which John Barrett, Fay Dowker, Renate
Loll and Andre Lukas presented very interesting but strikingly
different perspectives.  In the second round table, entitled {\it
Entropy and information: Classical \& quantum}, chaired by Joel
Lebowitz, John Baez spoke about entropy in the context of black hole
thermodynamics.  Finally, the congress had a Young Researchers
Symposium, with a number of plenary lectures intended to introduce
graduate students and post-docs to the exciting recent developments in
various areas of mathematical physics ranging from biophysics to
quantum chaos. I represented gravitational physics and spoke on the
{\it Interface of general relativity, quantum mechanics and
statistical physics.} All three sessions drew a large number of
participants also from other sub-fields of mathematical physics.

\vfill
\pagebreak

\section*{\centerline {3rd International LISA Symposium}}
\addcontentsline{toc}{subsubsection}{\it  
3rd International LISA Symposium, by Curt Cutler}
\begin{center}
    Curt Cutler, Max-Planck-Institut fuer Gravitationsphysik, Golm\\
\htmladdnormallink{
cutler@aei-potsdam.mpg.de}
{mailto:cutler@aei-potsdam.mpg.de}
\end{center}

The 3rd International LISA Symposium was held July 11-14, 2000 at the
Max-Planck-Institut fuer Gravitationsphysik in Golm, Germany. 
LISA Symposia are being held every two years, with venues
alternating between Europe and the United States. The first LISA
Symposium was held at RAL in July, 1996; the second 
was held in July, 1998 at Caltech. 
The main organizing bodies for the 3rd LISA Symposium 
were the Max-Planck-Institut fuer
Gravitationsphysik and the Max-Planck-Institut fuer Quantenoptik. 
The were about 100 participants. The Symposium proceedings will be
published as a special issue of Classical and Quantum Gravity 
in July, 2001. A detailed (320-page) description of the LISA Mission, the
recent LISA STS Report, is available
on-line at 
\htmladdnormallink{
ftp://ftp.rzg.mpg.de/pub/grav/lisa/sts\_1.02.pdf}
{ftp://ftp.rzg.mpg.de/pub/grav/lisa/sts\_1.02.pdf}

A LISA conference naturally begins with an update on 
the politics. Within ESA, LISA is 
already approved as a Horizon 2000+ Cornerstone Mission, 
but that status has little
practical worth, since ``approval'' leaves open 
the flight sequence, and without NASA cost-sharing
the flight probably could not happen before 2017. 
Practically, LISA's proponents within both NASA and ESA 
see LISA as a joint NASA/ESA mission to be flown around 2010.
There is now considerable enthusiasm for 
a shared LISA at NASA--so much enthusiasm that, 
as of this writing, Goddard is competing with
JPL over leadership of the project.
A cost-shared LISA would be a ``moderate mission'' from
NASA's perspective, lying within the SEU Program
(Structure and Evolution of the Universe).
It was  important that LISA did very well in the  recent
Taylor/McKee decadal review, ``Astronomy and Astrophysics in the
New Millennium''-- being the secondest-highest ranked 
``moderate'' mission. (GLAST was first.) 
The full report is at
\htmladdnormallink{
http://books.nap.edu/books/0309070317/html/}
{http://books.nap.edu/books/0309070317/html/}.

However the general perception is that, for LISA to get funded, 
there first needs to be technology demonstration mission, to 
be launched (one hopes) around 2005-6. 
Several possible avenues for this are being pursued;
as of this writing, the best bet seems to be 
a NASA ST3 mission, shared with ESA. 
The demonstrator mission would be a single satellite 
and would basically test the
drag-free system (which cannot be tested on the ground), with
the goal of demonstrating test mass isolation to
within $3\times 10^{-14} {\rm m s}^{-2} {\rm Hz}^{-1/2}$ 
between $1$ and $5$ mHz, i.e., within one order of magnitude
of the LISA goal.
(Noteworthy: two ``graybeards'' at the Symposium gave strong warnings
about the technology demonstrator. Rai Weiss warned repeatedly
that it should not be made too ambitious, since 1) you can't risk it
failing and 2) you don't want it to absorb all your time/energy.
Ben Lange, a pioneer of drag-free flight and
a veteran of many, many of successful space missions, 
advised that he'd ``avoid a demonstrator
mission like the plague.'') 
\vskip 0.20in
The Symposium included about 50 talks, which is too many to summarize. 
I'll confine myself to listing what were, to me, a few highlights,
and apologize in advance for the many excellent presentations I won't
even mention here, but which you'll be able to read in the Proceedings.

Sterl Phinney gave a beautiful and very upbeat summary of 
LISA sources.
New to me were the quite optimistic estimates 
for the merger rate for massive black hole binaries (MBH's), based
on a hierarchical clustering picture of structure formation, 
where small galaxies form first and merge to form bigger galaxies.
This picture leads to estimates of event rates of $\sim 0.1-10$/yr
for $10^6 M_\odot$ BH's out to $z=2$. But, importantly, LISA can
see far beyond $z=2$; Phinney argued that the merger rate
for  $\sim 10^5 M_\odot$ BH's out to $z \sim 20$ 
might be $\sim 1/$day.
\vskip 0.10in
John Armstrong and Massimo Tinto discussed their very important work (done
with Frank Estabrook), showing how one can (with some changes in hardware) 
linearly combine the LISA data
streams, with time delays, to form three linearly independent combinations
for which the laser phase noise exactly cancels. Two combinations contain
information on the two gw polarizations, and the third describes a 
``breathing mode'' that doesn't couple to GR. This third mode 
can be used to help calibrate and  
eliminate noise sources, and to discriminate between 
non-Gaussian noise bursts and real gw bursts.
\vskip 0.10in

There were several very interesting talks on
solar-mass compact objects spiraling
into MBH's. Scott Hughes showed that, when the 
MBH is near-extreme Kerr, the inspiral is strongly
affected by
superradiant scattering of gw's from the BH horizon.
Gw's that scatter off the horizon tend to ``hold up''
the test-body and increase the inspiral time by
$\sim 3\%$ (which is a lot of cycles).
Wolfgang Tichy discussed work-in-progress with E. Flanagan, 
claiming that, because the background Kerr metric is stationary, 
it actually {\it is} 
possible to determine how the Carter constant evolves from fluxes at infinity
(and at the horizon).
And Bernard Schutz discussed his worry 
(aroused, I assume, by recent work by Janna Levin) that because 
the orbits of spinning bodies in Kerr are chaotic, the number of matched 
filters will grow exponentially with the integration time, and
may be vastly greater than previously anticipated--effectively obliterating
the usual gains from matched filtering.
This was a warning, not  result, and somebody needs to look more 
carefully at this issue.
\vskip 0.10in
Large extra dimensions (perhaps as large as $0.3$ mm) 
are now much-discussed in string theory, and Craig Hogan
showed how these might lead to a gw background observable by LISA.
He argued that the early universe should produce 
copious gw's with wavelength comparable to the size
of the large extra dimension, which would be 
redshifted into the LISA band today.
Since the gw spectrum so-produced would be highly
peaked rather than flat, it is not constrained
by bounds at much lower frequencies coming 
from millisecond pulsar timing and COBE.
\vskip 0.10in
Ben Lange, who attended the whole meeting and then
gave us his outsider's perspective, made several recommendations,
especially emphasizing advantages of spherical test masses, instead
of cubes as in the current plan.
And he gave a delightful, short
summary of how in practice one can use a felt pen and
the classical mechanics of precession to
find the principal axes of an almost-perfect sphere. 
\vskip 0.10in Something new for a LISA Symposium: there were several
talks describing laboratory prototypes for LISA systems.  (Oliver
Jennrich: ``LISA is now more than just ink on paper.'')  Harry Ward
discussed his work on developing an interferometric read-out
system. He also reported on tests of how well hydrogen catalysis
bonding of optical elements would survive the rigors of launch and
space--and found the bonding held quite well under shaking and thermal
cycling.  Oliver Jennrich described his experiment showing the
feasibility of the LISA phase measurement scheme, using the same
amount of light as will be available for LISA.  Manuel Rodrigues
described a laboratory prototype for the inertial sensor, and Stefano
Vitale described a torsion pendulum test bench he is building to
testing the performance of the inertial sensor on the ground to some
$5\times 10^{-13} {\rm Newton}/\sqrt{\rm {Hz}}$. And Michael
Petersheim discussed a a prototype for the highly stable laser
required by LISA ($\Delta P/P < 4\times 10^{-4}/\sqrt{{\rm Hz}}$).

Lastly, there was very interesting discussion both of possible
variations in the LISA mission and possible follow-on missions (the
latter to be flown around 2020-2025, so fancy was free).  Bernard
Schutz pointed out the possible advantages of LISA starting out as
short-arm interferometer, before moving the satellites to the current
baseline separation of $5\times 10^6$ km.  He also suggested the
addition of a 4th spacecraft, to fly at the midpoint of one of the
three arms.  NASA has strongly encouraged LISA scientists to think
about possible follow-on missions to LISA, and is especially
interested in missions that might detect a primordial background of
gw's. There is little chance that LISA itself can detect a primordial
background, since in the LISA band it will be swamped by the
background from galactic and extra-galactic binaries.  The binary
background falls off at high frequencies, which leads to a
next-generation LISA concept featuring 3 constellations of
mini-LISA's, with the constellations forming a equilateral triangle
around the Sun at 1 AU, and each mini-LISA having short ($\sim
20,000$-km) arms to push the sensitivity band up to $1-10$ Hz. All of
this seems technologically feasible even in the near term--it's
``just'' a matter of money.

\end{document}